\documentclass[review]{elsarticle}

\usepackage{hyperref}

\journal{Signal Processing: Image Communication}

\usepackage[skip=2pt, font=small]{caption}
\usepackage{subcaption}
\usepackage{multirow}
\usepackage{graphicx}
\usepackage{amsmath,epsfig}
\usepackage{amssymb}

\begin{document}

\onecolumn 

\begin{enumerate}

\item[\textbf{Citation}]{Y. Hu, Z. Wang, and G. AlRegib, ``Texture classification using block intensity and gradient difference (BIGD),'' Signal Processing: Image Communication, vol. 83, 2020.}

\item[\textbf{DOI}]{\url{https://doi.org/10.1016/j.image.2019.115770}}

\item[\textbf{Review}]{Date of publication: April 2020}

\item[\textbf{Codes}]{\url{https://github.com/olivesgatech/}}

\item[\textbf{Bib}] {@article{hu2020bigd,\\
  title={Texture classification using block intensity and gradient difference (BIGD)},\\
  author={Hu, Yuting and Wang, Zhen and AlRegib, Ghassan},\\
  journal={Signal Processing: Image Communication},\\
  volume={83},\\
  year={2020},\\
  publisher={Elsevier}
}

}


\item[\textbf{Copyright}]{\textcopyright 2020 IEEE. Personal use of this material is permitted. Permission from IEEE must be obtained for all other uses, in any current or future media, including reprinting/republishing this material for advertising or promotional purposes,
creating new collective works, for resale or redistribution to servers or lists, or reuse of any copyrighted component
of this work in other works. }

\item[\textbf{Contact}]{\href{mailto:huyuting.sjtu@gmail.com}{huyuting.sjtu@gmail.com} OR \href{mailto:alregib@gatech.edu}{alregib@gatech.edu}\\
    \url{http://ghassanalregib.com/} \\ }

\end{enumerate}

\thispagestyle{empty}
\newpage
\clearpage
\setcounter{page}{1}

\begin{frontmatter}

\title{Texture Classification using Block Intensity and Gradient Difference (BIGD) Descriptor}

\author{Yuting Hu, Zhen Wang, and Ghassan AlRegib}
\address{Center for Signal and Information Processing\\
School of Electrical and Computer Engineering\\
Georgia Institute of Technology\\
Atlanta, GA 30332
}




\begin{abstract}
In this paper, we present an efficient and distinctive local descriptor, namely block intensity and gradient difference (\texttt{BIGD}). In an image patch, we randomly sample multi-scale block pairs and utilize the intensity and gradient differences of pairwise blocks to construct the local \texttt{BIGD} descriptor. The random sampling strategy and the multi-scale framework help \texttt{BIGD} descriptors capture the distinctive patterns of patches at different orientations and spatial granularity levels. We use vectors of locally aggregated descriptors (VLAD) or improved Fisher vector (IFV) to encode local \texttt{BIGD} descriptors into a full image descriptor, which is then fed into a linear support vector machine (SVM) classifier for texture classification. We compare the proposed descriptor with typical and state-of-the-art ones by evaluating their classification performance on five public texture data sets including Brodatz, CUReT, KTH-TIPS, and KTH-TIPS-2a and -2b. Experimental results show that the proposed \texttt{BIGD} descriptor with stronger discriminative power yields $0.12\%\sim6.43\%$ higher classification accuracy than the state-of-the-art texture descriptor, dense microblock difference (DMD).

\end{abstract}

\begin{keyword}
Local descriptor \sep block intensity and gradient difference (\texttt{BIGD}) \sep local feature extraction \sep multi-scale \sep texture classification
\end{keyword}

\end{frontmatter}


\section{Introduction}
\label{sec:intro}

Texture can be broadly defined as a type of visual features that characterize the surface of an object or a material. Distinctive and robust representation of texture is the key for various multimedia applications such as image representation~\cite{al2017curvelet}, texture retrieval~\cite{al2019curvelettextureretrieval}, face recognition\cite{zhao2017dynamic}, image quality assessment~\cite{temel2019perceptual,temel2016csv}, image/texture segmentation~\cite{vo2010study}, dynamic texture/scene recognition~\cite{arashloo2014dynamic,zhao2017dynamic}, texture/color style transfer~\cite{lo2016example}, and seismic interpretation\cite{long2018comparative}. Texture descriptors~\cite{bay2006surf,hu2016completed,liu2016evaluation, hu2017sselbp,liu2015fusing,mehta2016texture, allili2014texture}, which are robust against rotations and translations of images, are able to provide discriminative features. 

Texture representation requires texture descriptors to have two competing goals, high-quality description and low computational cost. Several research efforts have focused on extracting texture descriptors in a distinctive and efficient way. These approaches are commonly divided into several categories including covariance-, fractal-, filter-, gradient-, and binary-based descriptors. Covariance descriptors modeling the second-order statistics of images perform well on material categorization~\cite{harandi2014bregman}\cite{wang2012covariance}. However, by retaining only the second-order statistics, covariance descriptors are prone to be singular and have limited capability in modeling nonlinear, complicated feature relationships. Fractal-based descriptors~\cite{quan2014distinct,quan2014lacunarity} utilize the concept of fractal dimension and lacunarity analysis in fractal geometry to characterize the spatial distribution of local image structures in a statistical approach, guaranteeing both discriminative ability and invariance of texture descriptors. Filter-based descriptors~\cite{crosier2010using}\cite{varma2005statistical} acquire local features using filter banks. One problem of filter-based descriptors is that the design of filter banks is data dependent. Among gradient-based descriptors, scalable invariant feature transform (SIFT)~\cite{lowe2004distinctive} and speeded up robust features (SURF)~\cite{bay2006surf} as the most popular ones capture the discriminative gradient features of local patches. Binary descriptors such as local binary pattern (LBP)~\cite{ojala2002multiresolution}, binary robust independent elementary features (BRIEF)~\cite{calonder2010brief}, and local binary difference (LDB)~\cite{yang2014local}, which convert the intensity differences of neighboring pixels to binary values, are robust to monotonic illumination changes and require low computational cost. These advantages make binary descriptors more appealing for real-time applications. However, the binarization of local intensity differences leads to the loss of intensity information, which weakens the ability of discrimination.
Another disadvantage of binary descriptors is their dimensions, which will grow exponentially when the number of pairwise comparisons on neighboring pixels increases. To alleviate these problems, Zhang et al.~\cite{zhang2017feature} proposed a descriptor namely normalized difference vector (NDV) and Mehta et al.~\cite{mehta2016texture} proposed a novel descriptor called dense micro-block difference (DMD). Both methods composed of real-valued intensity differences instead of binary codes of different micro-blocks in local patches. Although DMD captures non-quantized patch-based features at multiple scales and orientations, the neglect of first-order gradients may deteriorate the discriminative ability~\cite{bay2006surf}\cite{yang2014local}.

In this paper, we introduce a novel local descriptor, block intensity and gradient difference (\texttt{BIGD}), which achieves great distinctiveness and computational efficiency. Compared with other algorithms mentioned above, our main contribution is efficiently capturing un-quantized gradient difference features in \texttt{BIGD}. The gradient difference captures the variations of gradients in a local patch and improves distinctiveness. Descriptors such as histogram of orientated gradients (HOG) and SIFT utilize gradient-based features to capture the orientation information. However, quantized orientations in them result in information loss. Our \texttt{BIGD} method extracts intensity- and gradient-difference features at multi-orientations without quantization and retains the discriminative power of features. 
To show the performance of \texttt{BIGD}, we evaluate it within a texture classification pipeline as shown in Fig.~\ref{fig:diagram}, which generally includes three major modules: a feature extraction, an image encoding, and a classifier. We utilize the our proposed descriptor in the feature extraction module. Since this paper mainly focuses on feature extraction rather than feature encoding and classifiers, we use simple feature encoding methods such as vectors of locally aggregated descriptors (VLAD)~\cite{jegou2010aggregating} or improved Fisher vectors (IFV)~\cite{perronnin2010improving}\cite{sanchez2013image} and classifiers such as a linear support vector machine (SVM) when we compare our proposed descriptor with other descriptors. The rest of this paper is organized as follows. Section~\ref{sec:BIGD_featureextraction} introduces the proposed \texttt{BIGD} descriptor and its application on texture classification. In Section~\ref{sec:experiments}, we evaluate the classification performance of the \texttt{BIGD} descriptor on five public texture databases and compare it with that of state-of-the-art methods. Section~\ref{sec:experiments} concludes the paper. Matlab codes will be available online\footnote{https://ghassanalregib.com/texturematerial-recognition-and-classfication/}.
\section{Proposed method}
\label{sec:BIGD_featureextraction}

\subsection{Block Pair Formulation}
\label{ssec:blkpair}

\begin{figure}[t]
\centering
\includegraphics[width=4.7in]{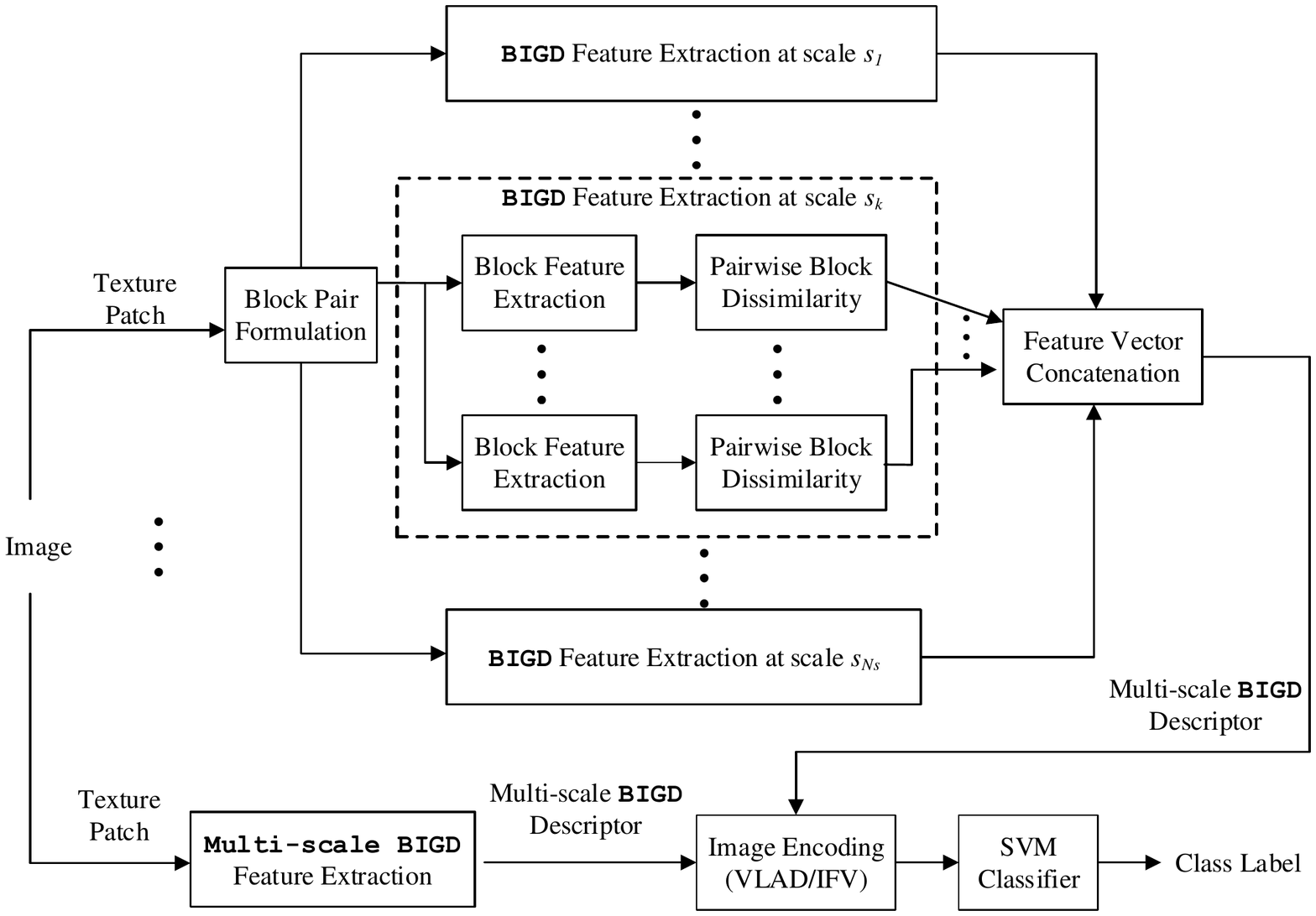}
\caption{The diagram of extracting multi-scale \texttt{BIGD} descriptors from a texture patch within a texture classification pipeline.}
\label{fig:diagram}
\end{figure}
\begin{figure}[t]
\centering
\includegraphics[width=2.5in]{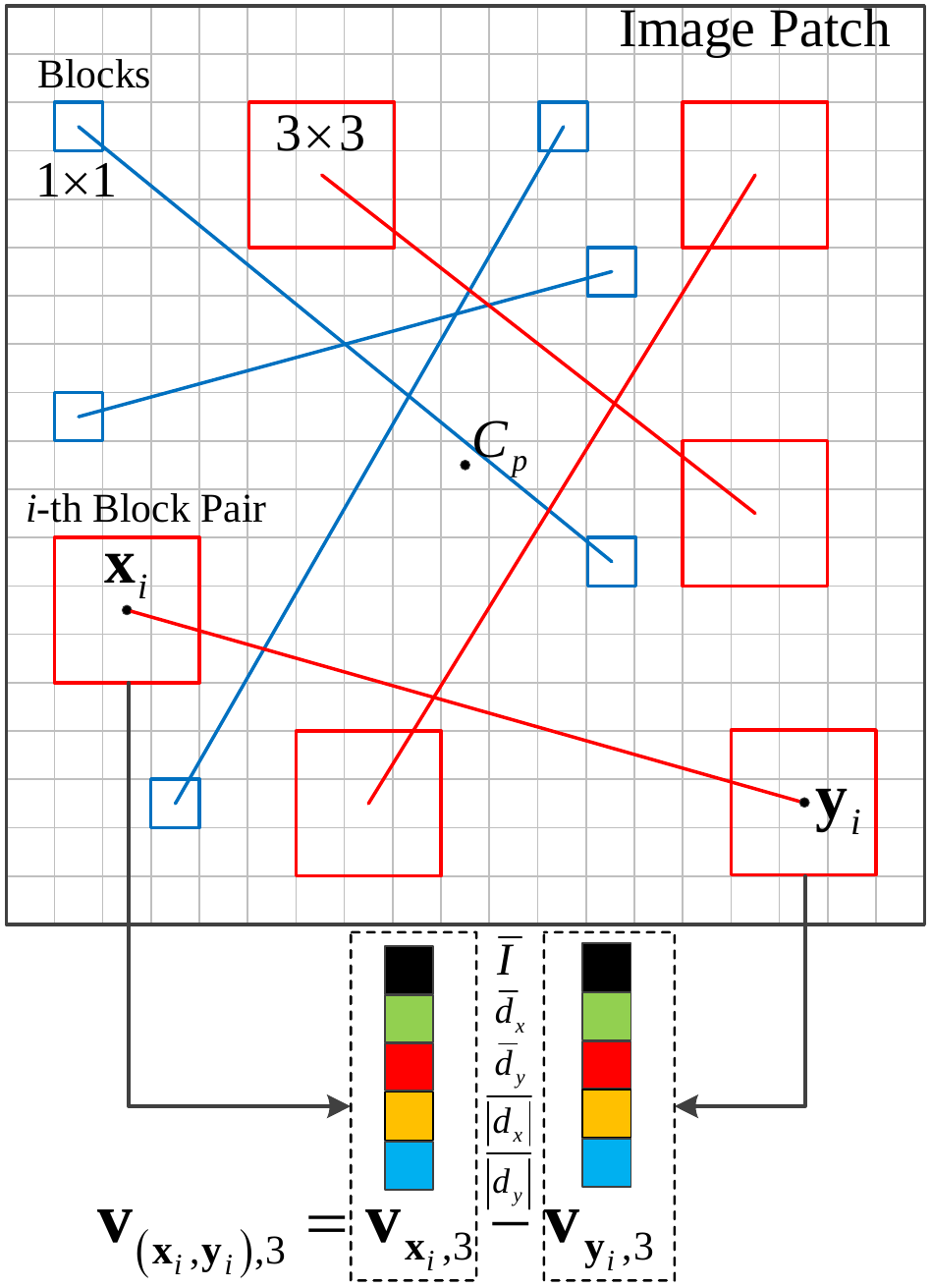}
\caption{Block pairs with different scales in an image patch and the feature difference of pairwise blocks $\left(\mathbf{x}_i, \mathbf{y}_i\right)$ at scale $3$.}
\label{fig:blkPs}
\end{figure}

The proposed \texttt{BIGD} describes the characteristic structures of patches that are evenly sampled with a step size of two pixels across the entire texture image and overlap with each other. The diagram that extracts the \texttt{BIGD} descriptor from a texture patch is shown in Fig.~\ref{fig:diagram}. To investigate the structural features of image patches, we randomly select multiple pairs of smaller square regions with various scales. Features extracted from these region pairs encode the local structures of patches at different spatial granularities and orientations and have higher robustness to noise than those extracted from raw pixels~\cite{mehta2016texture}. For simplicity, we specify these smaller square regions within the image patch as ``blocks''. An image patch of size $19\times 19$ centered at $C_p$ as an example in Fig.~\ref{fig:blkPs} contains block pairs connected by lines, where blue and red blocks have the sizes of $1\times 1$ and $3\times 3$, respectively. Only three block pairs at each scale are shown in Fig.~\ref{fig:blkPs}, but in our experiments we consider a greater number of block pairs (e.g. 4 block pairs/scale) at more scales (e.g. 4 scales). We denote block pairs as $\left(\mathbf{x}_i, \mathbf{y}_i\right)$, $i=1,2,\cdots, N$, where $N$ defines the number of block pairs in the image patch. As Fig.~\ref{fig:blkPs} shows, $\mathbf{x}_i=\left[x_{i1}, x_{i2}\right]$ and $\mathbf{y}_i=\left[y_{i1}, y_{i2}\right]$ are the coordinates of the central pixels of the two blocks belonging to the $i$-th pair. Since blocks in the image patch are randomly selected, we identify the centers of all pairwise blocks using two sets of sampling points, $X=\left\{\mathbf{x}_1,\mathbf{x}_2,\cdots,\mathbf{x}_N\right\}$ and $Y=\left\{\mathbf{y}_1,\mathbf{y}_2,\cdots,\mathbf{y}_N\right\}$. In the image patch, the coordinates of all block centers are represented by the coordinate system with the origin at patch center $C_p$. Following the sampling strategy in~\cite{calonder2010brief}, we select sampling points in $X$ and $Y$ from the isotropic Gaussian distribution, denoted $\left(X, Y\right)\thicksim \mbox{i.i.d.}$ Gaussian$(0, L^2/25)$, where $L$ is the size of image patches. $\left(X, Y\right)$ define the positions of $N$ block pairs relative to an image patch.
During the process of feature extraction, we will keep these block pairs unchanged for all patches over the entire image.

\subsection{Multi-scale \texttt{BIGD} Descriptor Extraction}
\label{sec:bigd}

\begin{figure*}[t]
\centering
\includegraphics[width=4.8in]{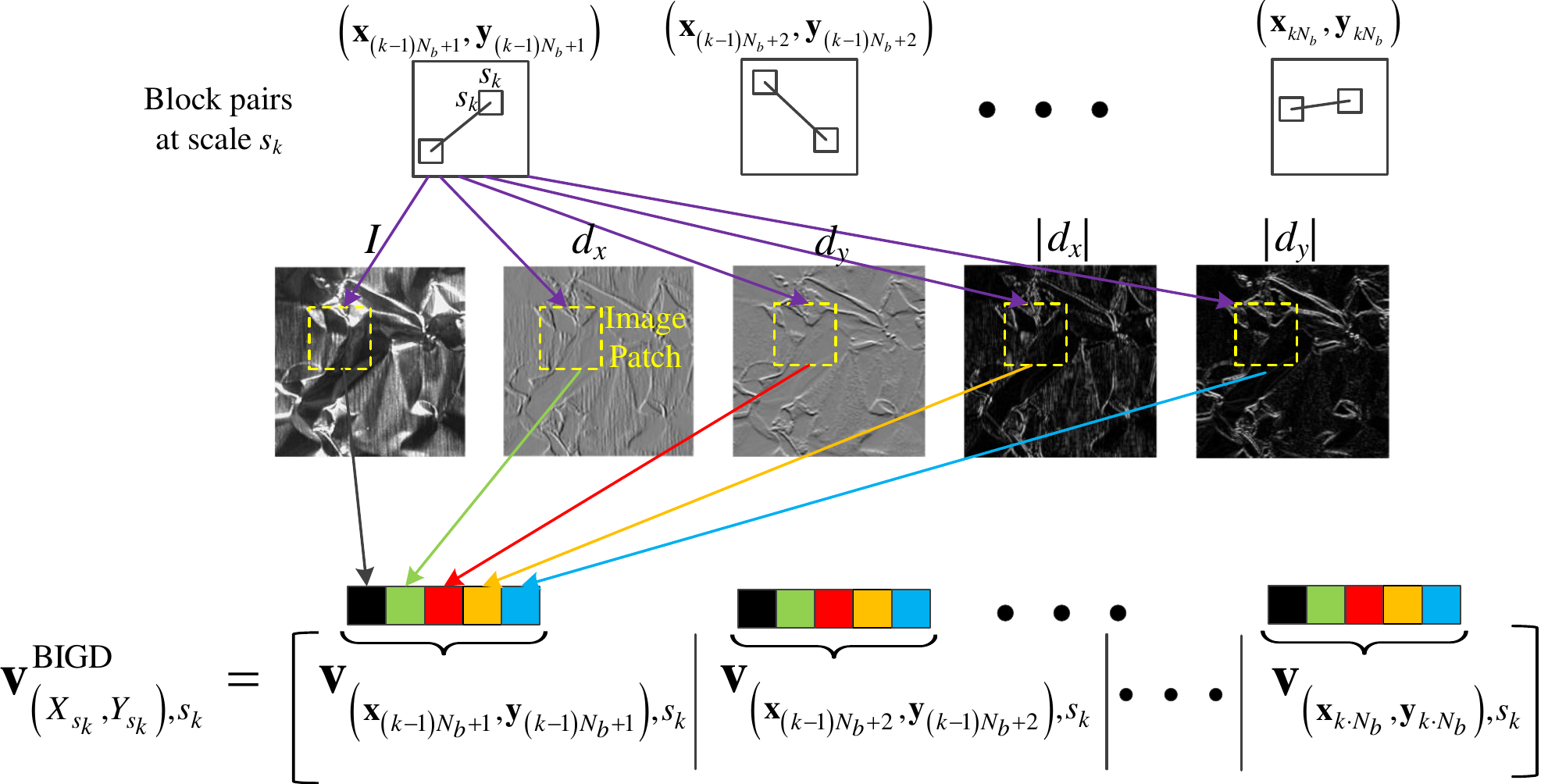}
\caption{The extraction of the \texttt{BIGD} descriptor from an image patch using randomly sampled block pairs at scale $s_k$.}
\label{fig:BIGD}
\end{figure*}

\subsubsection{\texttt{BIGD} feature extraction}
\label{sssec:bigd_feature}
\begin{figure}[t]
\begin{minipage}[b]{0.32\linewidth}
  \centering
  \centerline{\includegraphics[width=2.5cm]{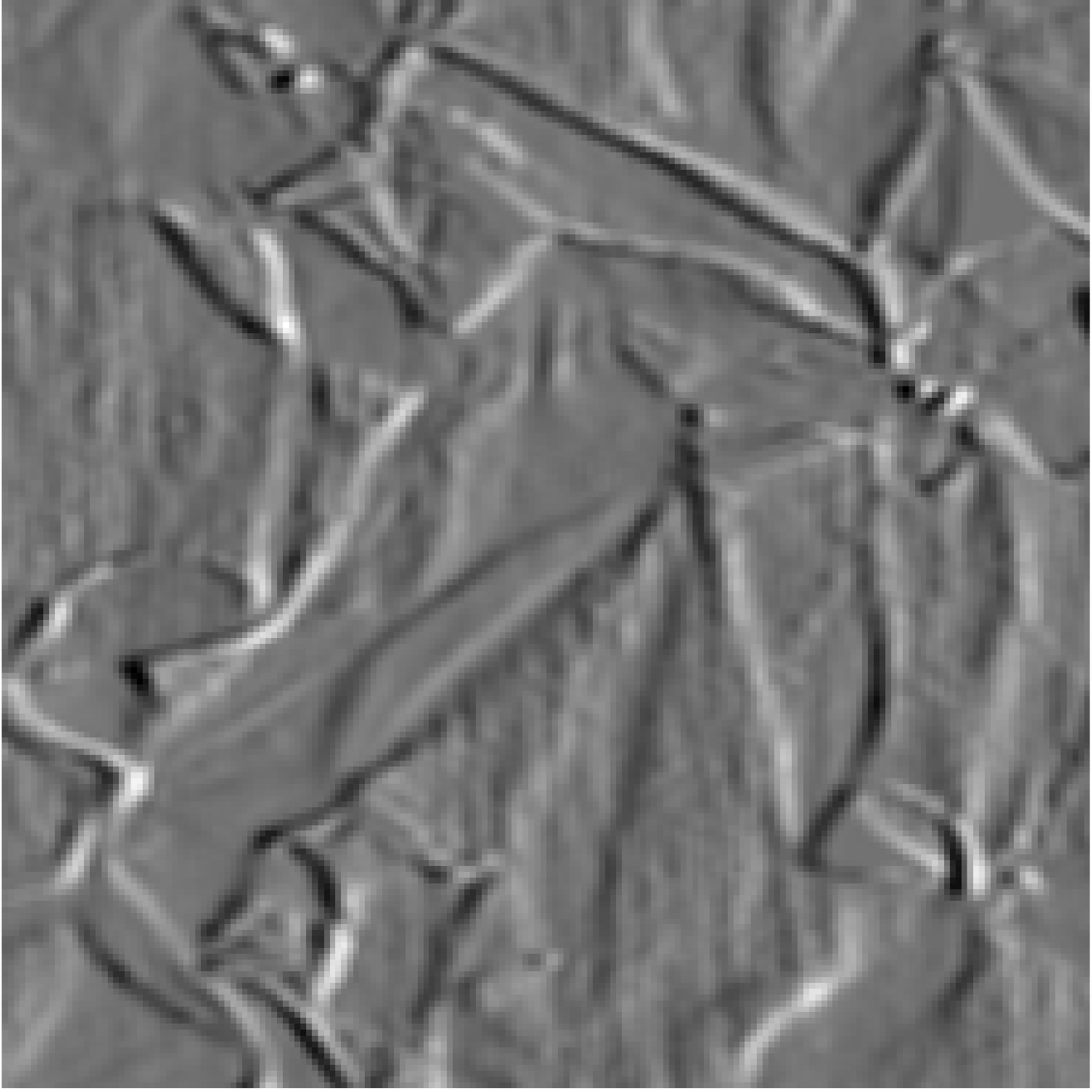}}
  \centerline{(a) $\bar{I}$}\medskip
\end{minipage}
\begin{minipage}[b]{0.32\linewidth}
  \centering
  \centerline{\includegraphics[width=2.5cm]{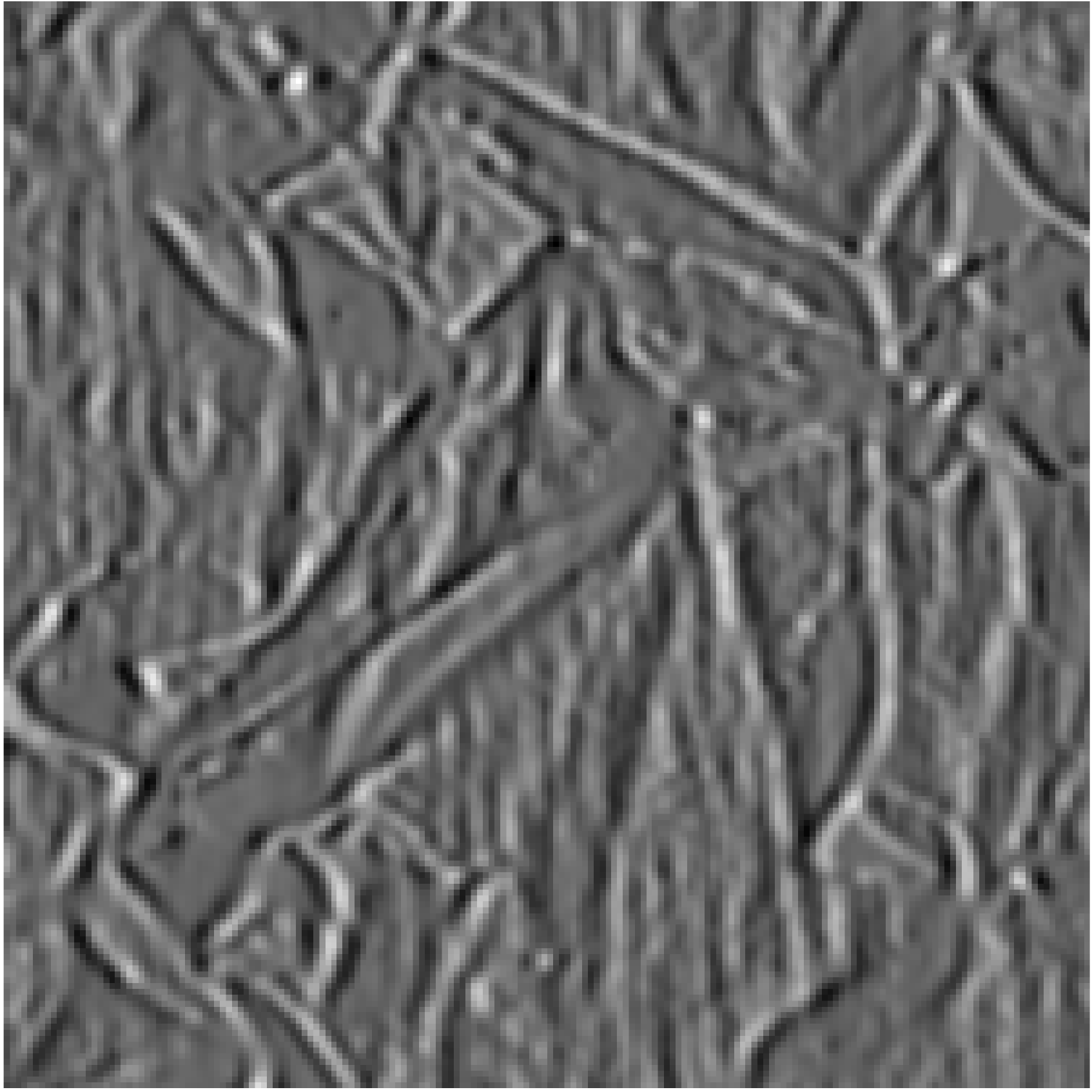}}
  \centerline{(b) $\bar{d}_x$}\medskip
\end{minipage}
\begin{minipage}[b]{0.32\linewidth}
  \centering
  \centerline{\includegraphics[width=2.5cm]{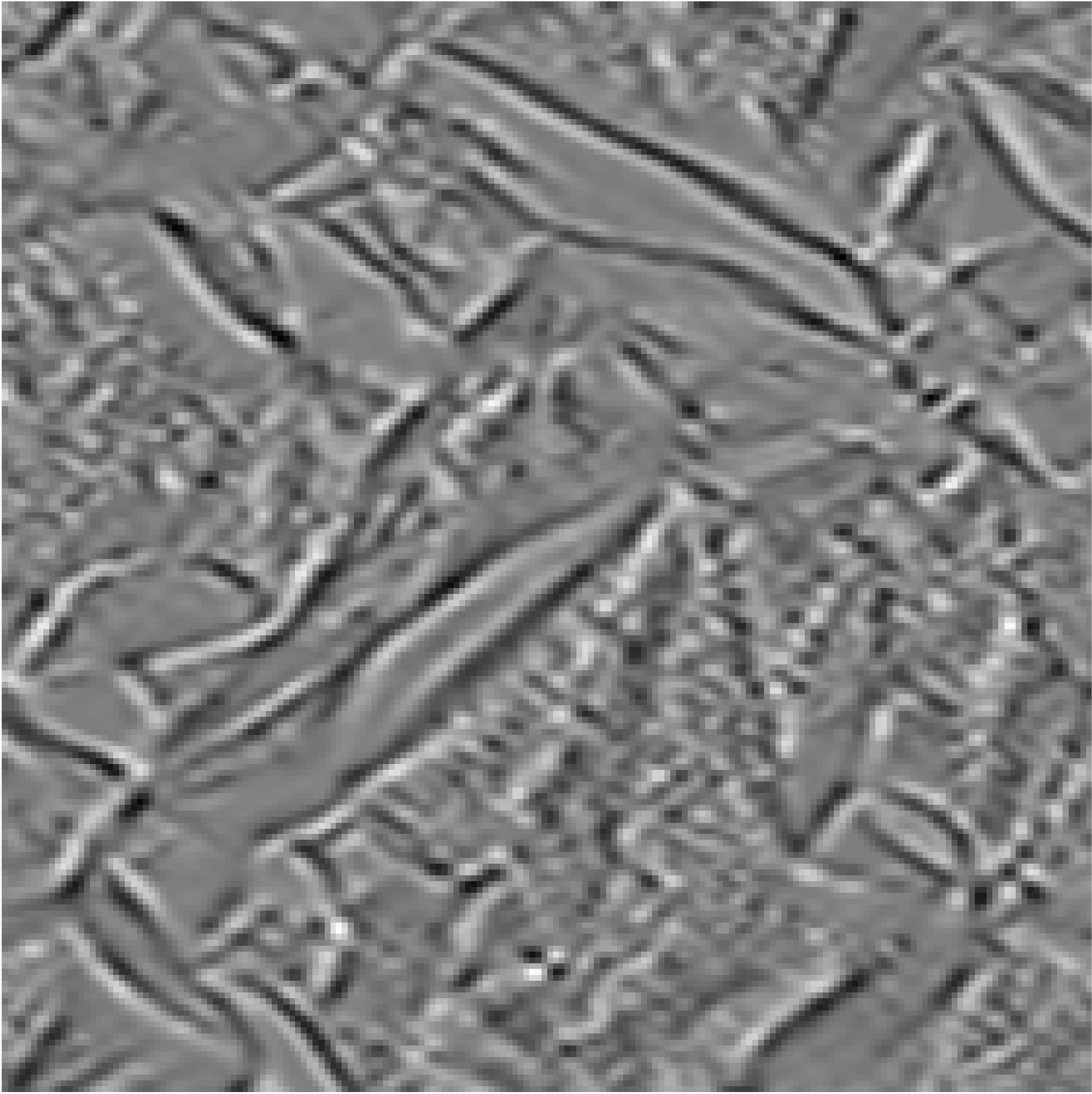}}
  \centerline{(c) $\bar{d}_y$}\medskip
\end{minipage}
\\
\begin{minipage}[b]{0.45\linewidth}
  \centering
  \centerline{\includegraphics[width=2.5cm]{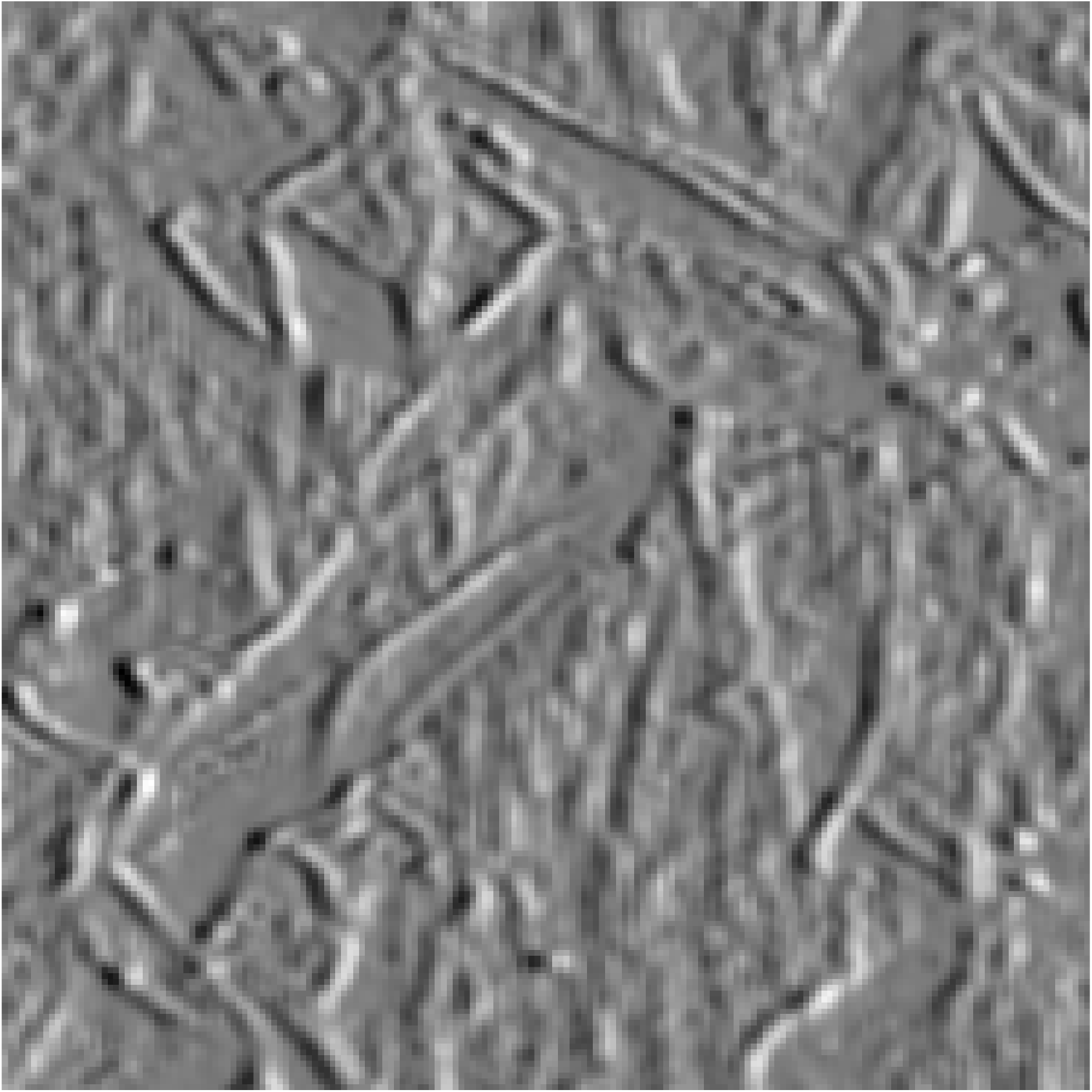}}
  \centerline{(d) $\overline{|d_x|}$}\medskip
\end{minipage}
\begin{minipage}[b]{0.45\linewidth}
  \centering
  \centerline{\includegraphics[width=2.5cm]{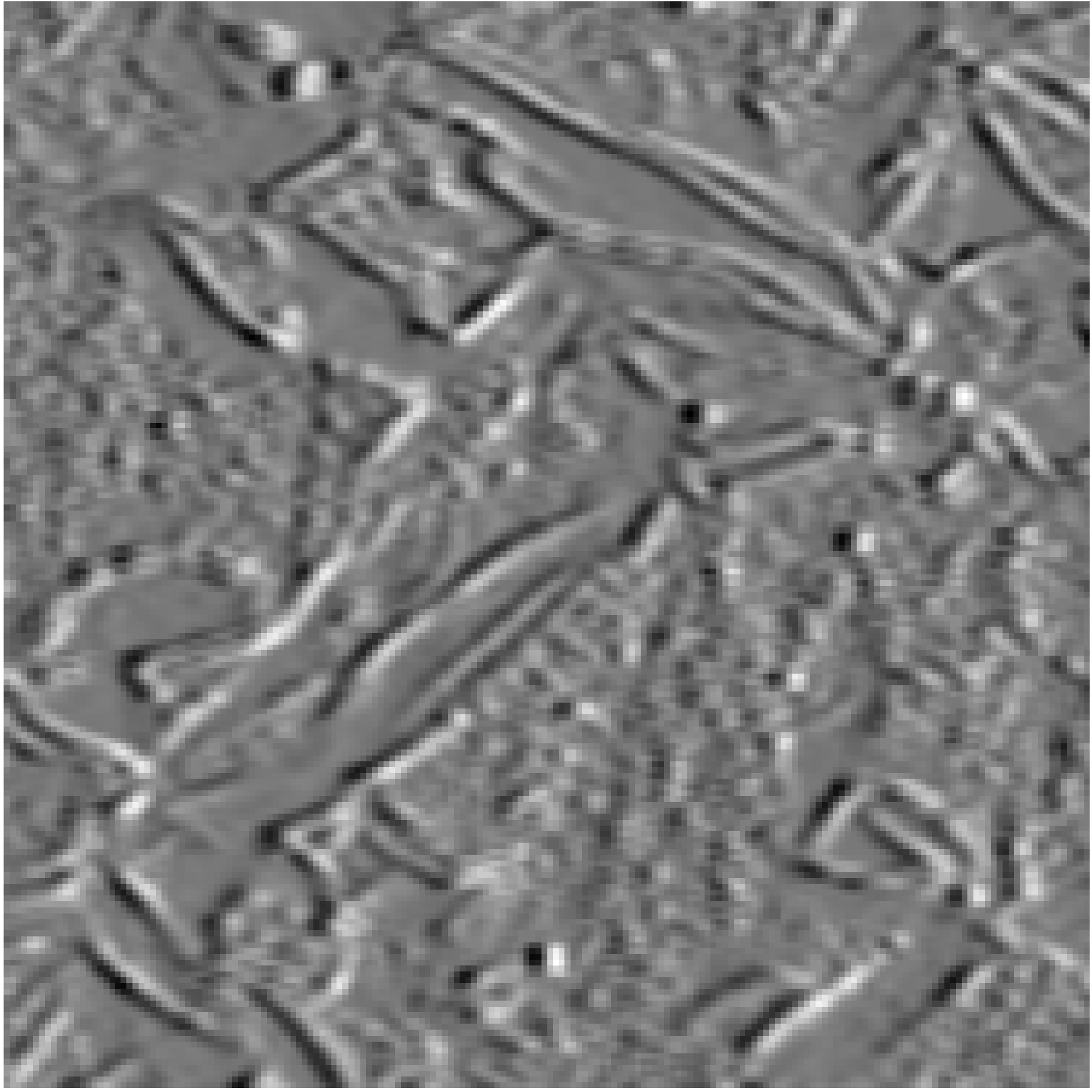}}
  \centerline{(e) $\overline{|d_y|}$}\medskip
\end{minipage}
\caption{Five feature maps extracted by block pairs $\left([-2,-5],[-2,-1]\right)$ at scale 3 from all patches of an image.}
\label{fig:featuremaps}
\end{figure}

By comparing the difference between pairwise blocks in various perspectives, we describe the local structures of patches. As introduced in~\cite{mehta2016texture}, the average intensity difference of pairwise blocks captures variations in an image patch. Here, we denote the average intensity of blocks as $\bar{I}$. However, depending only on this feature, we cannot properly characterize the dissimilarity of pairwise blocks. Therefore, we propose to utilize the average horizontal and vertical gradients of blocks, denoted $\bar{d}_x$ and $\bar{d}_y$, respectively, to capture smoothness. To obtain $\bar{d}_x$ and $\bar{d}_y$, we first apply the Sobel operator on all pixels in the patch and then average the horizontal and vertical gradients of pixels in blocks. In addition, in order to analyze the polarity of intensity changes in patches, we average the absolute values of horizontal and vertical gradients and obtain another two features, denoted $\overline{|d_x|}$ and $\overline{|d_y|}$, respectively. Therefore, the block centered at $\mathbf{x}_i$ with scale $s$ corresponds to a five-dimensional feature vector, denoted $\mathbf{v}_{\mathbf{x}_i,s}=\left(\bar{I}, \bar{d}_x, \bar{d}_y, \overline{|d_x|}, \overline{|d_y|}\right)$. The dissimilarity between pairwise blocks $\left(\mathbf{x}_i, \mathbf{y}_i\right)$ at scale $s$ is evaluated by the difference between the feature vectors of corresponding blocks, denoted $\mathbf{v}_{\left(\mathbf{x}_i, \mathbf{y}_i\right), s}=\mathbf{v}_{\mathbf{x}_i,s}-\mathbf{v}_{\mathbf{y}_i,s}$.
For clarity, we define $\mathbf{v}_{\left(\mathbf{x}_i, \mathbf{y}_i\right), s}$ as the feature vector of pairwise block $\left(\mathbf{x}_i, \mathbf{y}_i\right)$ at scale $s$. The bottom of Fig.~\ref{fig:blkPs} shows an example of calculating $\mathbf{v}_{\left(\mathbf{x}_i, \mathbf{y}_i\right), 3}$ in a patch, where color squares represent different features. After extracting $\mathbf{v}_{\left(\mathbf{x}_i, \mathbf{y}_i\right), s}$ from all patches, we obtain five feature maps. Fig.~\ref{fig:BIGD} shows the process of extracting the \texttt{BIGD} descriptor of an image patch using randomly sampled block pairs at scale $s_k$, where the first row illustrates the random sampling strategy. The second row of Fig.~\ref{fig:BIGD} represents raw intensity and gradient maps, from which $\mathbf{v}_{\left(\mathbf{x}_i, \mathbf{y}_i\right), s_k}$ is calculated as shown in the third row of Fig.~\ref{fig:BIGD}. In Fig.~\ref{fig:featuremaps}, we use block pair $(\mathbf{x}_i, \mathbf{y}_i)$ at scale 3, where $\mathbf{x}_i=[-2,-5]$ and $\mathbf{y}_i=[-2,-1]$, to obtain five feature maps from raw intensity and gradient maps shown in Fig.~\ref{fig:BIGD}. Notably these feature maps contain similar structures and that is because all features are extracted from the same block pair. However, different details in feature maps provide more information about the local structures of patches.

\subsubsection{Multi-scale extraction scheme}
\label{sssec:multiscale}
The random selection of block pairs determines that extracted features can describe the local structure of patches in various orientations.
To acquire a more discriminative representation of patches, we sample block pairs at multiple scales. We denote a set of scales as $S=\{s_1,s_2,\cdots,s_{N_s}\}$, where $N_s$ represents the number of scales. Since an image patch contains $N$ block pairs and we assume that every scale is of the same importance, the number of block pairs at each scale is $N_b=N/{N_s}$. We rewrite $X$ and $Y$ in Section~\ref{ssec:blkpair} as $X=\left[X_{s_1}, X_{s_2},\cdots, X_{s_{N_s}}\right]$ and $Y=\left[Y_{s_1}, Y_{s_2},\cdots, Y_{s_{N_s}}\right]$, respectively, to identify block pairs at different scales. $\left(X_{s_k}, Y_{s_k}\right)$, $k=1,2,\cdots, N_s$, which contains the centers of pairwise blocks at scale $s_k$, can be expressed as follows:

\begin{equation}
\label{equ:XYsk}
\begin{aligned}
\left(X_{s_k}, Y_{s_k}\right)=&\left\{\left(\mathbf{x}_{(k-1)N_b+1},\mathbf{y}_{(k-1)N_b+1}\right),\left(\mathbf{x}_{(k-1)N_b+2},\right.\right.\\
&\ \ \ \left.\left.\mathbf{y}_{(k-1)N_b+2}\right),\cdots,\left(\mathbf{x}_{k N_b},\mathbf{y}_{kN_b}\right)\right\}.
\end{aligned}
\end{equation}

On the basis of $(X_{s_k}, Y_{s_k})$, in an image patch we calculate the features of pairwise blocks at scale $s_k$ and concatenate feature vectors to generate the corresponding \texttt{BIGD} descriptor at scale $s_k$, denoted $\mathbf{v}_{(X_{s_k},Y_{s_k}),s_k}^{\textnormal{BIGD}}$, as follows:
\begin{equation}
\label{equ:bigd}
\begin{aligned}
\mathbf{v}_{(X_{s_k},Y_{s_k}),s_k}^{\textnormal{BIGD}}=\left[\mathbf{v}_{(\mathbf{x}_{(k-1)N_b+1},\mathbf{y}_{(k-1)N_b+1}),
s_k}|\cdots|\mathbf{v}_{(\mathbf{x}_{kN_b},\mathbf{y}_{kN_b}),s_k}\right],
\end{aligned}
\end{equation}
Fig.~\ref{fig:BIGD} illustrates the process that extracts the \texttt{BIGD} descriptor at scale $s_k$. By concatenating \texttt{BIGD} descriptors at all scales, we obtain the \texttt{BIGD} descriptor at all scales, denoted $\mathbf{v}_{(X,Y),S}^{\textnormal{BIGD}}$, which describes the local structures of an image patch at different granularities and orientations. The expression of $\mathbf{v}_{(X,Y),S}^{\textnormal{BIGD}}$ is shown as follows:
\begin{equation}
\label{equ:multibigd}
\mathbf{v}_{(X,Y),S}^{\textnormal{BIGD}}=\left[\mathbf{v}_{(X_{s_1},Y_{s_1}),s_1}^{\textnormal{BIGD}}, \cdots, \mathbf{v}_{(X_{s_{N_s}},Y_{s_{N_s}}),s_{N_s}}^{\textnormal{BIGD}}\right].
\end{equation}

\begin{figure*}[t]
\centering
\begin{minipage}[b]{0.45\linewidth}
  \centering
  \centerline{\includegraphics[height=4.4cm]{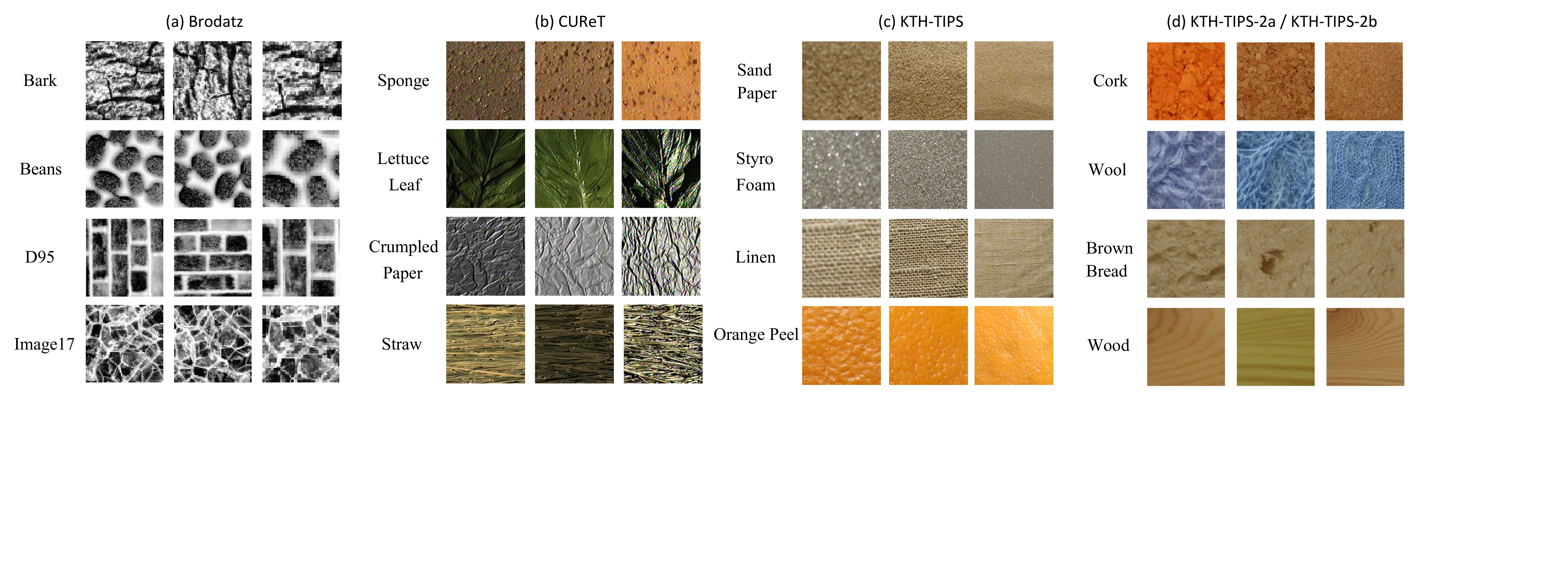}}
  \centerline{(a) Brodatz}\medskip
\end{minipage}
\hfill
\begin{minipage}[b]{0.45\linewidth}
  \centering
  \centerline{\includegraphics[height=4.4cm]{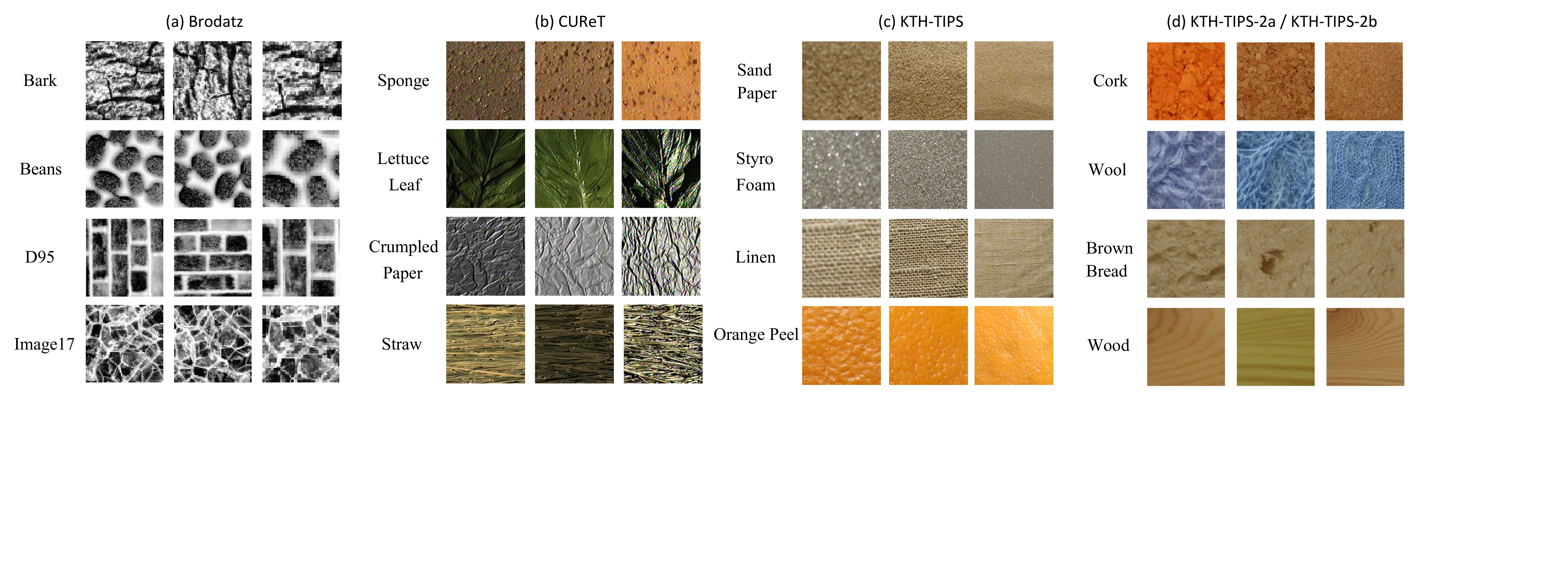}}
  \centerline{(b) CUReT}\medskip
\end{minipage}
\hfill
\begin{minipage}[b]{0.45\linewidth}
  \centering
  \centerline{\includegraphics[height=4.4cm]{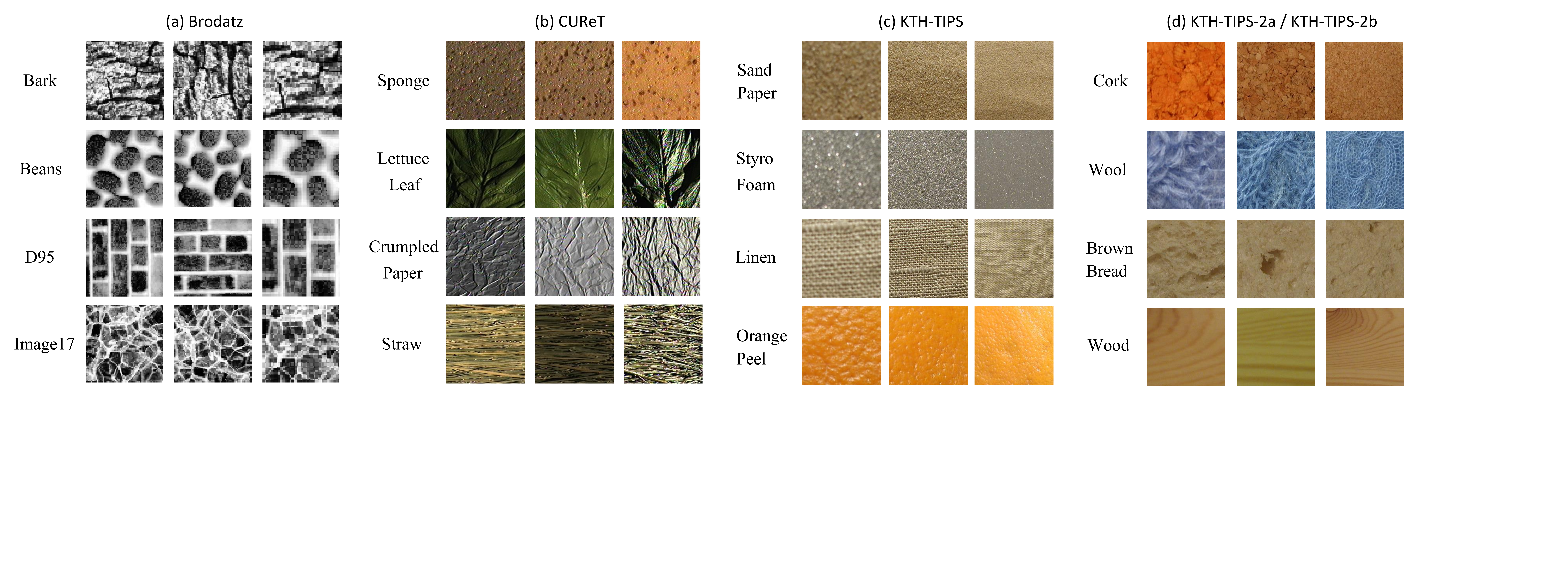}}
  \centerline{(c) KTH-TIPS}\medskip
\end{minipage}
\hfill
\begin{minipage}[b]{0.45\linewidth}
  \centering
  \centerline{\includegraphics[height=4.4cm]{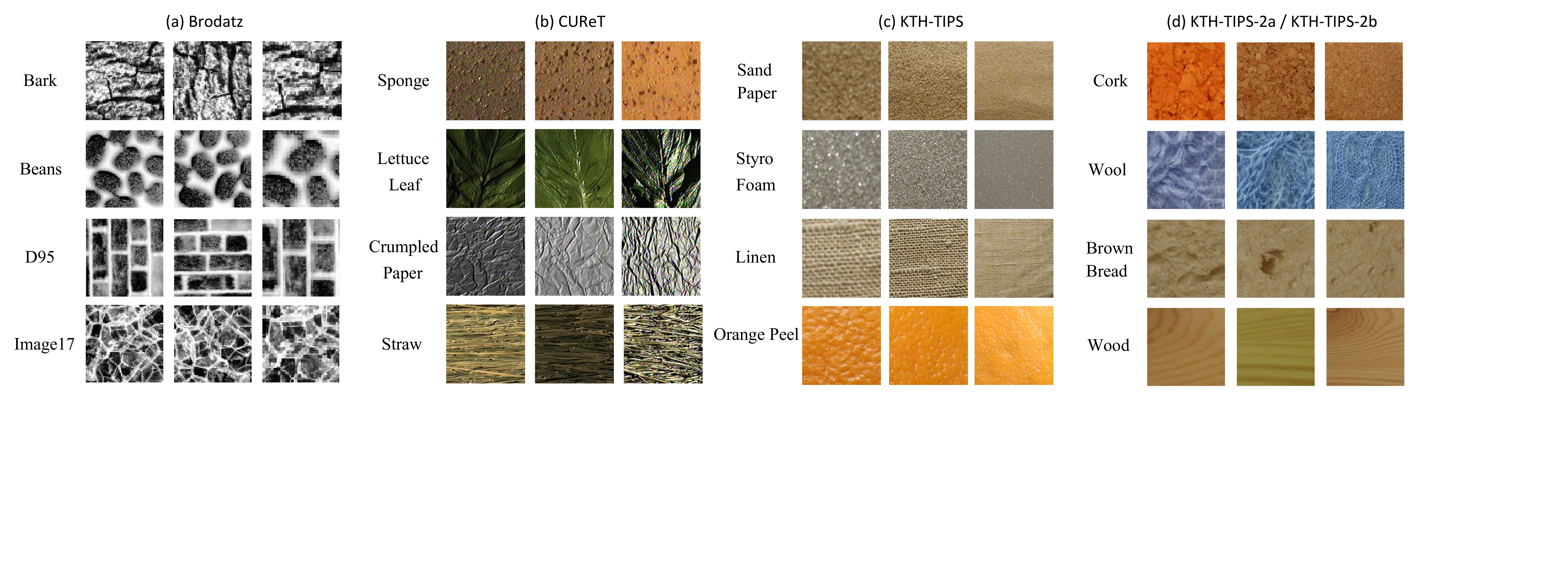}}
  \centerline{(d) KTH-TIPS-2a/-2b}\medskip
  \label{fig:database:KTH-TIPS-2ab}
\end{minipage}
\caption{Five typical public databases for texture classification: Brodatz, CUReT, KTH-TIPS, and KTH-TIPS-2a/-2b.}
\label{fig:database}
\end{figure*}
\begin{table*}[t]
    \begin{center}
    \caption{Descriptions of five public texture databases} \label{tab:database}
    \resizebox{1.0\textwidth}{!}{
        \begin{tabular}{|c|c|c|c|c|c|c|l|}
          \hline
          \multirow{2}{*}{\centering Databases }    &    $\#$    & \# & Image Size & $\#$ Images & $\#$ Train & $\#$  Test  & \multirow{2}{*}{\centering Capturing Conditions}\\
             & Classes  & Images & (pixels) & /Class & /Class & /Class  & \\
          \hline
          Brodatz~\cite{chen2010wld}      & 32     &2048      & $200\times200$      &       64            &        32          &        32          & 16 samples, rotation- and scale- variations\\
          \hline
          CUReT~\cite{varma2009statistical}        & 61      & 5612     & $200\times200$      &       92            &        46          &        46          & 1 sample, various viewing angles and illuminants\\
          \hline
          KTH-TIPS~\cite{mallikarjuna2006kth}    & 10   &810        & $200\times200$      &        81           &        40          &          41        & 1 sample, 3 viewing angles, 3 illuminants, and 9 scales      \\
          \hline
          KTH-TIPS-2a~\cite{mallikarjuna2006kth} & 11  &4752         & $200\times200$      &        432          &        324         &      108           & 4 samples, 3 viewing angles, 4 illuminants, and 9 scales \\
          \hline
          KTH-TIPS-2b~\cite{mallikarjuna2006kth}  & 11  &4752           & $200\times200$      &        432          &        324         &      108           & 4 samples, 3 viewing angles, 4 illuminants, and 9 scales \\
          \hline
        \end{tabular}
        }
    \end{center}
\end{table*}

\subsection{Image Encoding and Classification}
\label{sec:encoding}

 High-dimensional feature representation is suitable for the combination use with linear SVM. Since VLAD encoding and FV encoding extract high-dimensional features, we evaluate both of them. FV extends the BOW by encoding higher order statistics (first and second order) while VLAD accumulates the differences of local features assigned to each codeword. Though VLAD is a simplified version of FV, it is differential and easily generalized to residual layers in the design of convolutional neural networks.

\subsubsection{VLAD encoding}
\label{secsec:vlad}

To encode the \texttt{BIGD} descriptors of all patches into a full image descriptor, we utilize a typical encoding method, VLAD~\cite{jegou2010aggregating}, which is the simplified form of the Fisher vector (FV). Following the conventional notations of VLAD encoding, we denote \texttt{BIGD} descriptor $\mathbf{v}_{(X,Y),S}^{\textnormal{BIGD}}$ as $x\in\mathbb{R}^d$, where $d$ represents the dimension of $\mathbf{v}_{(X,Y),S}^{\textnormal{BIGD}}$. We first partition \texttt{BIGD} descriptors extracted from the patches of training images into $K$ clusters using k-means clustering~\cite{leung2001representing}. K-means clustering is a commonly used unsupervised vector quantization method for learning a codebook of visual words (i.e. textons) and it aims to partition feature vectors into $K$ clusters in which each feature vector belongs to the cluster with the nearest mean. The corresponding cluster centers, denoted $\left\{u_i\right\}_{i=1}^K$, $u_i\in\mathbb{R}^d$, as codewords, construct a codebook. Then from each image, we assume that we extract $m$ \texttt{BIGD} descriptors, denoted $\chi=\left\{x_t\right\}_{t=1}^m$, $x_t\in\mathbb{R}^d$. By finding the closest codeword to $x_t$, we partition $\left\{x_t\right\}_{t=1}^m$ into $K$ groups.
In each group, we obtain vector $v_i$ by accumulating differences between codeword $u_i$ and its corresponding \texttt{BIGD} descriptors. The expression of $v_i$ is shown as follows:
\begin{equation}
\label{equ:localdifferencevector}
v_i = \sum_{x_t\in u_i} \left(u_i-x_t\right).
\end{equation}
Finally, by concatenating $\left\{v_i\right\}_{i=1}^{K}$, we obtain the encoded descriptor of an image with the length of $dK$. 

\subsubsection{IFV encoding}
\label{secsec:FV}
Fisher encoding~\cite{perronnin2010improving} uses Gaussian mixture models (GMM) to represent the distribution of local \texttt{BIGD} descriptors and captures the derivatives of GMM with respect to model parameters. Given prior probability $\pi_k$, mean $u_k$, and covariance matrix $\sum_k$, $k=\{1,2, \cdots, K\}$, denoted $\Theta = \{ \pi_k, \mu_k, \Sigma_k; k \in \{1,2, \cdots, K\}\}$, the distribution of \texttt{BIGD} descriptor $x\in \mathbb{R}^d$ can be described by $p(x|\Theta)=\sum_{k=1}^{K}\pi_k p \left(x|u_k,\Sigma_k\right)$.
To learn model parameters $\pi_k$, $u_k$, and $\sum_k$, we apply expectation maximization (EM) to \texttt{BIGD} descriptors extracted from the patches of training images. Then we calculate the derivatives of $\log p(x|\Theta)$ with respect to $u_k$ and $\sum_k$ as follows:
\begin{equation}
\label{equ:fv1}
\left\{
\begin{aligned}
&\frac{\partial \log p(x|\Theta)}{\partial \mu_{k}}
        =h_k \Sigma_{k}^{-1}(x-\mu_k)\\
&\frac{\partial \log p(x|\Theta)}{\partial \Sigma_{k}^{-1}}
    =\frac{h_k}{2} \left( \Sigma_{k}-(x-\mu_k)^2\right )
\end{aligned}
\right. ,
\end{equation}
where $h_k=\frac{\pi_k p \left(x|u_k,\Sigma_k\right)}{\sum_k \pi_k p \left(x|u_k,\Sigma_k\right)}$. FV encoding concatenates all derivatives for the K components of GMM and obtains a vector with the length of $2dK$. The details of FV encoding can be found in~\cite{perronnin2010improving}, and in our experiments we use its improved version, IFV~\cite{sanchez2013image} because of its better representation ability than FV shown in~\cite{sanchez2013image}, which uses signed-square-root embedding followed by $L_2$ normalization.

\subsubsection{SVM classifier}
\label{secsec:SVM}
To categorize texture images, we apply a classifier on image descriptors encoded by VLAD or IFV. Given labeled training data, a SVM~\cite{cortes1995support} classifier outputs an optimal hyperplane in a multi-dimensional space to separate different classes. SVM classifiers, which have been proved to have better performance on texture classification~\cite{liu2015fusing}, have two main advantages, less training time and direct operations on features.  Therefore, we feed image descriptors into SVM classifiers with a linear kernel. Because of the simple structure of SVM classifiers, we can attribute the improvement of classification performance to extracted features rather than classifiers.

\section{Experimental Evaluations}
\label{sec:experiments}

\subsection{Database}
\label{secsec:database}
In order to show the superiority of the proposed \texttt{BIGD} descriptor over other state-of-the-art texture descriptors, we are going to evaluate their corresponding performance on texture classification. In this section, we conduct a set of experiments on five public texture databases: Brodatz~\cite{chen2010wld}, CUReT~\cite{varma2009statistical}, KTH-TIPS~\cite{mallikarjuna2006kth}, and KTH-TIPS-2a and -2b~\cite{mallikarjuna2006kth}, in which texture images are captured under various conditions with the changes of occlusions, viewpoints, and illuminants. 
To illustrate the changes of capturing conditions, we randomly select four classes of textures from each database and exhibit three samples of every selected class in a row as Fig.~\ref{fig:database} shows. Texture images from KTH-TIPS-2a and -2b are shown in Fig.~\ref{fig:database}(d) together since these two databases have the same texture classes. 
Although we notice that some samples are color images, in our experiments we use only gray-scale images. To keep consistency and ensure fair comparison with~\cite{mehta2016texture}, all images have the same size $200 \times 200$. For the Brodatz dataset, we apply a bilinear or bicubic interpolation to resize the image from the original size $64 \times 64$ to $200 \times 200$. For other datasets, we can download images with $200 \times 200$ from their official websites. We give a brief description of each database in Table~\ref{tab:database}, which involves the number of classes, the number of total images, the image size, the number of images per class, the numbers of training and testing images per class, and capturing conditions. The combinations of various capturing conditions on image samples generate all texture images in each class. For example, although each class of the KTH-TIPS database has only one image sample, from this image sample the combinations of capturing conditions including three viewing angles, three illuminants, and nine scales generate $3\times3\times9=81$ images. In addition, the numbers of training and testing images per class determine the testing protocol of texture classification. 

The original Brodatz database~\cite{valkealahti1998reduced} consists of 32 classes of textures, each of which contains 16 image samples. By applying rotation, scaling, or both operations on original image samples, we obtain an extended database, in which each class contains $16\times 4=64$ images. Following the testing protocol in~\cite{chen2010wld}, for each class, we randomly select half of images (i.e., 32 images/class) for training and use the remaining (i.e., 32 images/class) for testing.

In contrast to the extended Brodatz database, the CUReT database~\cite{varma2009statistical} is more challenging for texture classification. It consists of images acquired under various viewing angles and illuminants, which result in significant changes of texture appearances. The CUReT database is composed of 61 classes, each of which contains 92 images. The testing protocol in \cite{liu2016evaluation} requires us to randomly select half of images in each class (i.e., 46 images/class) for training and use the remaining (i.e., 46 images/class) for testing. 

As an extension of the CUReT database, the KTH-TIPS~\cite{mallikarjuna2006kth} database selects a subset of images from the CUReT database and adds scale variations to these images. The KTH-TIPS database consists of ten classes, and each class contains 81 images captured under three viewing angles, three illuminants, and nine scales. According to the testing protocol in~\cite{liu2016evaluation}, in each class we randomly select half of images (i.e., 40 images/class) for training and use the remaining (i.e., 41 images/class) for testing.

As the two extensions of the KTH-TIPS database, KTH-TIPS-2a and -2b databases, which are designed for the recognition of surface materials, contain the images of 11 classes of materials such as wood and wool. In these two databases, each class consists of four physical samples, and each physical sample corresponds to 108 images captured under three viewing angles, four illuminants, and nine scales. Following the test protocol in~\cite{liu2016evaluation}, we use three physical samples of each class (i.e., 324 images/class) for training and the remaining one (i.e., 108 images/class) for testing.

\subsection{Implementation Details}
\label{ssec:implementation}
The general pipeline of texture classification consists of three main modules: feature extraction, image encoding, and classification. In the first module, texture descriptors focus on describing the representative features of texture images. To obtain the \texttt{BIGD} descriptor that efficiently captures structural details of texture patches, first of all, we evenly sample the centers of patches with a step size of two pixels across the entire texture image. Every sampled center corresponds to a local patch with a size of $L\times L = 15\times 15$. For local patches, we apply a Gaussian random sampling strategy and select $N=16$ block pairs. In our experiments, the local patches of all images in a database share the same layout of block pairs. To generate the more discriminative representations of patches, we extract \texttt{BIGD} descriptors in a multi-scale framework, where pairwise blocks have $N_s=4$ \mbox{scales} ranging from $1\times 1$ to $4\times 4$. Under the assumption that each scale is of same importance, the number of block pairs at each scale is $N_b=N/N_s=4$. From a block pair at one scale, we extract five features involving intensity and gradient differences. 
Therefore, by concatenating the feature vectors of block pairs at all scales, we obtain the multi-scale \texttt{BIGD} descriptor of a local patch with a dimension of $d=5N_bN_s=80$.

Image coding as the second module of texture classification encodes local \texttt{BIGD} descriptors into a full image descriptor using VLAD or IFV. An important step of image coding is to obtain model parameters trained on the local \texttt{BIGD} descriptors of training images. For example, the KTH-TIPS database have $400$ (i.e., $40$ training images/class $\times 10$ classes) training images with the size of $200\times 200$. By sampling patch centers with a step of two in each training image, we identify $10,000$ patches. Rather than using all $4,000,000$ patches sampled from training images, in practical implementation we randomly select $500,000$ patches for computational efficiency. By training the local descriptors of selected patches, we obtain the codebook of VLAD and parameter set $\Theta$ of IFV. We set the number of clusters $K$ as $128$ in both VLAD and IFV for consistency and use the MATLAB\textsuperscript{\textregistered} VLFeat toolbox~\cite{Vedaldi10a} to implement k-means, GMM, VLAD, IFV encoding, and SVM. Since we used ``vl\_kmeans'' in the MATLAB\textsuperscript{\textregistered} VLFeat toolbox to implement k-means, its default setting is heuristic ``Lloyd" algorithm for k-means clustering. Regarding the learning parameter settings for SVM such as regularization parameter (e.g. $\lambda = 1/(\#classes\times \#training\ images)$ and maximum number of iterations (e.g. $100\times $\#training\ images), we use the same setting as DMD~\cite{mehta2016texture} for a fair comparison. To guarantee fair comparisons between the proposed method and state-of-the-art ones, we keep the parameter setting unchanged for all databases unless we specify it.

In the SVM classification module, we randomly split each database into training and testing sets using testing protocols in Table~\ref{tab:database} and repeat the partition ten times. In the tables of this paper, we use two metrics, the average and standard deviation of classification accuracy over ten splits, to evaluate the performance of various descriptors on texture classification.

\subsection{Effects of Parameters}
\label{secsec:parameters}
The performance of \texttt{BIGD} descriptors on texture classification depends on several factors such as patch sizes, block sizes, the number of k-means or GMM clusters, and the testing protocol. 
To understand the effects of these parameters on the performance of texture classification, we conduct our experiments mainly on Brodatz and KTH-TIPS-2a databases. These two databases are selected because of the great difference between their corresponding texture types. 
For simplicity, experiments in this section sample block pairs at a fixed scale and utilize VLAD as the encoding method. 
In addition, testing protocols will follow Table~\ref{tab:database} unless we specify it. 

\subsubsection{Patch and block sizes}
According to our previous discussion, patch size $L\times L$ and block size $s\times s$ determine local \texttt{BIGD} descriptors. Therefore, we extract local \texttt{BIGD} descriptors with various parameter pairs $(L,s)$ from Brodatz and KTH-TIPS-2a databases and list the corresponding classification results in Tables~\ref{tab:Brodatz_windowsize} and~\ref{tab:KTH-TIPS-2a_windowsize}, respectively. In these two tables, we notice that if the block size is fixed, the increasing of the patch size improves average classification accuracy. The main reason is that patches with a larger size cover more local details and provide more choices of block pairs without involving redundant information, which comes from the overlapping of block pairs. In Table~\ref{tab:Brodatz_windowsize}, for the Brodatz dataset, the proposed method using parameter pair $(L, s)=(15, 3)$ achieves the highest classification accuracy $99.8\%$. In contrast, as Table~\ref{tab:KTH-TIPS-2a_windowsize} shows, for the KTH-TIPS-2a database, the proposed method with parameter pair $(L, s) = (13, 2)$ has the classification accuracy of $83.23\%$, which achieves at most a $4.64\%$ increase compared to other parameter pairs. The best choice of parameter pair $(L, s)$ changes with databases, which implies that fixed patch and block sizes may not be able to accurately capture the details of textures. Therefore, to generate the more universal and discriminative representations of patches in different databases, we sample block pairs at multiple scales when extracting local \texttt{BIGD} descriptors. Coarse-level blocks reduce the effect of noise while fine-level blocks capture the details of local patterns. In addition, according to Sec.~\ref{sec:bigd}, the dimension of local \texttt{BIGD} descriptors keeps unchanged regardless of the scales of block pairs. Therefore, the comparison between classification accuracy in Tables~\ref{tab:Brodatz_windowsize} and~\ref{tab:KTH-TIPS-2a_windowsize} and that of the proposed multi-scale \texttt{BIGD} descriptor is meaningful. 
\begin{table}[t]
    \begin{center}
    \caption{Classification accuracy of the proposed method on the Brodatz database using different $(L,s)$ pairs.}
    \label{tab:Brodatz_windowsize}

    \resizebox{0.8\textwidth}{!}{
        \begin{tabular}{|c|c|c|c|c|}
          \hline
         Block Size & \multicolumn{4}{|c|}{Patch Size ($L\times L$)}\\
          \cline{2-5}
          ($s\times s$) & $9 \times 9$ & $11 \times 11$ & $13 \times 13$ & $15 \times 15$\\
          \hline
          $1 \times 1$ & $99.29 \pm 0.24$      &  $99.59 \pm 0.16$     &   $99.64 \pm 0.25$              &    $99.71 \pm 0.25$                    \\
          \hline
          $2 \times 2$ &$99.56 \pm 0.30$       & $99.60 \pm 0.25$      &   $99.69 \pm 0.22$              &  $99.78 \pm 0.15$                       \\
          \hline
          $3 \times 3$ &  $99.53 \pm 0.22$     & $99.63 \pm 0.30$      & $99.71 \pm 0.19$                & $\mathbf{99.80}\pm 0.20$                        \\
          \hline
          $4 \times 4$ &  $99.55 \pm 0.21$      & $99.49 \pm 0.27$      &  $99.70 \pm 0.17$               &  $99.68 \pm 0.29$                        \\
          \hline
          $5 \times 5$ &   $99.32 \pm 0.49$    & $99.62 \pm 0.28$      &  $99.61 \pm 0.21$               &   $99.59 \pm 0.26$                        \\
          \hline
        \end{tabular}
    }
    \end{center}
\end{table}
\begin{table}[t]
    \begin{center}
    \caption{Classification accuracy of the proposed method on the KTH-TIPS-2a database using different $(L,s)$ pairs.}
    \label{tab:KTH-TIPS-2a_windowsize}
    \resizebox{0.9\textwidth}{!}{
        \begin{tabular}{|c|c|c|c|c|}
          \hline
         Block Size & \multicolumn{4}{|c|}{Patch Size ($L\times L$)}\\
          \cline{2-5}
          ($s\times s$) & $9 \times 9$ & $11 \times 11$ & $13 \times 13$ & $15 \times 15$\\
          \hline
          $1 \times 1$ & $80.46 \pm 3.48$      &  $79.71 \pm 3.09$     &   $80.27 \pm 2.85$              &    $82.86 \pm 1.83$                    \\
          \hline
          $2 \times 2$ &$80.74 \pm 3.92$       & $79.91 \pm 2.93$      &   $\mathbf{83.23} \pm 3.64$              &  $82.10 \pm 2.80$                       \\
          \hline
          $3 \times 3$ &  $80.24 \pm 4.89$     & $81.47 \pm 3.68$      & $81.53 \pm 4.02$                & $80.50 \pm 3.52$                        \\
          \hline
          $4 \times 4$ &  $79.16\pm 6.18$      & $80.79 \pm 3.65$      &  $79.21 \pm 3.75$               &  $80.65 \pm 4.44$                        \\
          \hline
          $5 \times 5$ &   $78.59 \pm 2.50$    & $80.66 \pm 2.65$      &  $81.52 \pm 3.66$               &   $80.38 \pm 3.54$ \\
          \hline
        \end{tabular}
    }
    \end{center}
\end{table}


\subsubsection{Numbers of k-means clusters}
K-means clusters describe the distribution of local feature descriptors and the number of clusters effect the encoding performance. To explore the effect of the number of k-means clusters, $K$, we select a variety of $K$ values ranging from $16$ to $128$ and list their corresponding classification accuracy on Brodatz and KTH-TIPS-2a databases in Table~\ref{tab:numberofkmeans}. For consistency, we set parameter pair $(L, s)$ for the Brodatz  database as $(15, 3)$ and for the KTH-TIPS-2a database as $(13, 2)$, which correspond to the best classification performance in Tables~\ref{tab:Brodatz_windowsize} and~\ref{tab:KTH-TIPS-2a_windowsize}. As we mentioned above, if the dimension of a local \texttt{BIGD} descriptor is $d$, a full image descriptor encoded by VLAD has a dimension of $dK$. A higher $K$ value corresponds to a more abundant vocabularies but leads to the increasing of feature dimensionality. In Table~\ref{tab:numberofkmeans}, we notice that the classification accuracy of two datasets keeps growing with the increase of $K$. Although $K$ with a value larger than $128$ may correspond to higher classification accuracy, for computational efficiency, we set $K$ as $128$ for all experiments unless mentioned otherwise.

\begin{table}[t]
    \begin{center}
    \caption{Classification accuracy of the proposed method on Brodatz and KTH-TIPS-2a databases using the different numbers of k-means clusters.}
    \label{tab:numberofkmeans}
    \resizebox{1.0\textwidth}{!}{
        \begin{tabular}{|c|c|c|c|c|c|}
          \hline
          Cluster & \multirow{2}*{$16$} & \multirow{2}*{$32$} & \multirow{2}*{$64$} &  \multirow{2}*{$96$} &  \multirow{2}*{$128$}\\ 
          Numbers($K$) &  &  &  & &\\
          \hline
          Brodatz & $98.47 \pm 0.42$ & $99.41 \pm 0.30$ & $99.62 \pm 0.22$ & $99.69 \pm 0.22$ & $\mathbf{99.80}\pm 0.20$\\
          \hline
          KTH-TIPS-2a & $77.25 \pm 3.26$ &  $81.51 \pm 1.86$ &   $82.93 \pm 3.75$ &    $80.54 \pm 1.80$ & $\mathbf{83.23} \pm 3.64$  \\
          \hline
        \end{tabular}
    }
    \end{center}
\end{table}

\begin{table}[t]
    \begin{center}
    \caption{Classification accuracy of the proposed method on Brodatz and KTH-TIPS-2a databases using different testing protocols.}
    \label{tab:numberoftrains}
    \resizebox{0.85\textwidth}{!}{
        \begin{tabular}{|c|c|c|c|}
          \hline
          Training \# vs. Testing \# & $1:3$ & $1:1$ & $3:1$\\
          \hline
          Brodatz & $98.46 \pm 1.11$ & $99.80 \pm 0.20$ & $99.80 \pm 0.18$ \\
          \hline
          KTH-TIPS-2a  & $67.35 \pm 2.57$  & $76.39 \pm 2.43$ & $83.23 \pm 3.64$\\
          \hline
        \end{tabular}
    }
    \end{center}
\end{table}
\subsubsection{Testing protocols}
In addition to several factors mentioned above, the testing protocol or the ratio between the numbers of training and testing images also has an effect on classification performance. The changes of classification accuracy under different testing protocols reflect the robustness of the proposed method. Following the testing protocol in~\cite{mehta2016texture}, we test our proposed approach on Brodatz and KTH-TIPS-2a databases by choosing parameter pairs $(L,s)$ same to Table~\ref{tab:numberofkmeans} and setting the number of k-means clusters as $128$. The classification results of three testing protocols on two databases are shown in Table~\ref{tab:numberoftrains}. It is certain that more training images lead to higher classification accuracy and smaller standard deviations. In addition, we notice that even though the ratio between the numbers of training and testing images is $1:3$, our proposed method is still able to achieve the classification accuracy of $98.46\%$ for the Brodatz database and $67.35\%$ for the KTH-TIPS-2a database. This supports our claim that the combination of \texttt{BIGD} descriptors and VLAD has strong potentials on the discrimination of texture images.

\begin{table*}[htb!]
    \begin{center}
    \caption{Performance comparison between the proposed method and other typical and state-of-the-art methods}
    \label{tab:stateoftheart}
        \begin{subtable}[t]{.45\linewidth}
            \footnotesize
            \caption{Brodatz}   \label{tab:stateoftheart_Brodatz}
           \resizebox{\textwidth}{!}{
            \begin{tabular}{|c|c|}
              \hline
               \multirow{2}*{Methods} & Classification\\
                                      & Accuracy ($\%$) \\
                \hline
                LBP\cite{ojala2002multiresolution} &       $87.2$                       \\
              \hline
              LQP\cite{ul2012visual}
              &   $96.9$                             \\
              \hline
              WLD\cite{chen2010wld}
                    &  $96.5$                             \\
               \hline
               LHS\cite{sharma2012local}
             &    99.30                           \\
              \hline
              SIFT+IFV\cite{mehta2016texture} &$97.6$ \\
              \hline
              SDMD+IFV\cite{mehta2016rotation}
                    &  $99.7$                             \\
              \hline
              DMD+IFV$^{\ast}$\cite{mehta2016texture}  & $99.8\pm0.2$                       \\
              \hline
              \hline
              \texttt{BIGD}+IFV           & $\mathbf{99.9}\pm 0.1$ \\
              \hline
              \texttt{BIGD}+VLAD           &   $99.7\pm0.1 $ \\
              \hline
            \end{tabular}
        }
        \end{subtable}
        \begin{subtable}[t]{0.35\linewidth}
            \footnotesize
            \caption{CUReT}
            \label{tab:stateoftheart_curet}
            \resizebox{\textwidth}{!}{
            \begin{tabular}{|c|c|}
              \hline
               \multirow{2}*{Methods} & Classification\\
                                      & Accuracy ($\%$) \\
                \hline
                MR8\cite{varma2005statistical}      &    $93.5$                            \\
              \hline
              BIF\cite{crosier2010using}      &    $95.8$                            \\
              \hline
              RP\cite{liu2012texture}      &    $98.5 $                           \\
               \hline
              CLBP\cite{liu2016evaluation}      &    $97.3$                            \\
              \hline
              SIFT+IFV\cite{mehta2016texture}     &    $98.1$                            \\
              \hline
              DMD+IFV$^{\ast}$\cite{mehta2016texture}  &    $98.4 \pm 0.7$                     \\
              \hline
              \hline
              \texttt{BIGD}+IFV           & $ \mathbf{99.0} \pm  0.5$    \\
              \hline
              \texttt{BIGD}+VLAD           & $98.1 \pm 0.9$    \\
              \hline
              \hline
              FV-AlexNet\cite{cimpoi2015deep}  & $98.4$    \\
              \hline
              FV-VGGM\cite{cimpoi2015deep} & $98.7$    \\
              \hline
              FV-VGGVD\cite{cimpoi2015deep}   & $99.0$    \\
              \hline
              DeCAF\cite{cimpoi2014describing} & $97.9 \pm 0.4$\\
              \hline
              DeCAF+IFV\cite{cimpoi2014describing} & $99.8 \pm 0.1$\\
              \hline
            \end{tabular}
        }
        \end{subtable}
        \begin{subtable}[t]{.32\linewidth}
            \footnotesize
            \caption{KTH-TIPS}
            \label{tab:stateoftheart_kthtips}
            \resizebox{\textwidth}{!}{
            \begin{tabular}{|c|c|}
              \hline
               \multirow{2}{*}{Methods} & Classification\\
                                      & Accuracy ($\%$) \\
               \hline
               SSELBP\cite{hu2017sselbp} & $98.1$ \\
              \hline
              BIF\cite{crosier2010using} & $98.5$ \\
              \hline
              SRP\cite{liu2015fusing} & $\mathbf{99.3}$ \\
              \hline
              COV-LBPD\cite{hong2014combining} & $98.0$\\
               \hline
              WLD\cite{chen2010wld} & $91.1$\\
 
              \hline
 SIFT+IFV\cite{mehta2016texture}     &    $97.3$ \\
              \hline
              DMD+IFV$^{\ast}$\cite{mehta2016texture} & $97.6 \pm 1.6$\\
              \hline
              \hline
              \texttt{BIGD}+IFV & $98.8 \pm 1.1$\\
              \hline
              \texttt{BIGD}+VLAD & $\mathbf{99.0} \pm 0.8$ \\
              \hline
              \hline
              DeCAF\cite{cimpoi2014describing}  &    $97.0 \pm 0.9$                     \\
              \hline  
              DeCAF+IFV\cite{cimpoi2014describing}  &    $99.8 \pm 0.2$                     \\
              \hline  
            \end{tabular}
        }
        \end{subtable}
        \begin{subtable}[t]{.32\linewidth}
            \footnotesize
            \caption{KTH-TIPS-2a}
            \label{tab:stateoftheart_kthtips2a}
            \resizebox{\textwidth}{!}{
            \begin{tabular}{|c|c|}
              \hline
               \multirow{2}*{Methods} & Classification\\
                                      & Accuracy ($\%$) \\
              \hline
              MS-BIF\cite{crosier2010using}          &    $71.6$                 \\
              \hline
              LHS\cite{sharma2012local}              &    $73.0$                  \\
              \hline
              COV-LBPD\cite{hong2014combining}       &    $74.9$                  \\
              \hline
              scLBP\cite{ryu2015sorted}              &    $78.4$                   \\
              \hline
             
             NDV\cite{zhang2017feature} &    $77.1$               \\
              \hline
              SIFT+IFV\cite{mehta2016texture} &    $76.6$               \\
              \hline
              DMD+IFV$^{\ast}$\cite{mehta2016texture}    &    $80.3 \pm 6.1$          \\
              \hline
              \hline
              \texttt{BIGD}+IFV                       & $\mathbf{81.3} \pm 3.6$     \\
              \hline
              \texttt{BIGD}+VLAD             & $81.2 \pm 2.5$    \\
              \hline
              \hline
             DeCAF\cite{cimpoi2014describing}  &    $78.4 \pm 2.0$                     \\
              \hline DeCAF+IFV\cite{cimpoi2014describing}  &    $84.7 \pm 1.5$                     \\
              \hline
            \end{tabular}
        }
        \end{subtable}
        \begin{subtable}[t]{.34\linewidth}
            \footnotesize
            \caption{KTH-TIPS-2b}
            \label{tab:stateoftheart_kthtips2b}
            \resizebox{\textwidth}{!}{
            \begin{tabular}{|c|c|}
              \hline
               \multirow{2}*{Methods} & Classification\\
                                      & Accuracy ($\%$) \\
               \hline
              MC\_SBP\cite{nguyen2016statistical}  &    $71.6$                     \\
              \hline
              CDL\cite{wang2012covariance}  &    $76.3$                     \\
              \hline
              S$_{H}$-SVM\cite{harandi2014bregman}  &    $80.1$                     \\
              \hline
             Timofte\cite{timofte2012training}              &    $66.3$                 \\
              \hline
              DMD+IFV$^{\ast}$\cite{mehta2016texture}  &    $76.2 \pm 4.1$                     \\
              \hline
              \hline
              \texttt{BIGD}+IFV                       &       $ 81.4     \pm   3.1$   \\
              \hline
              \texttt{BIGD}+VLAD              & $\mathbf{82.7} \pm 4.5$    \\
              \hline
              \hline
               FV-AlexNet\cite{cimpoi2015deep}    & $77.9$    \\
              \hline
               FV-VGGM\cite{cimpoi2015deep}      & $79.9$    \\
              \hline
               FV-VGGVD\cite{cimpoi2015deep}     & $88.2$    \\
              \hline
            \end{tabular}
        }
        \end{subtable}
    \end{center}
\end{table*}

\subsection{Experimental Results}
\label{secsec:results}

We run experiments on Matlab and the code will be made available. In order to compare the classification performance of the proposed method with those of typical and state-of-the-art ones, we conduct our experiments on five public texture databases, Brodatz, CUReT, KTH-TIPS, and KTH-TIPS-2a and -2b by following standard testing protocols listed in Table~\ref{tab:database}. The parameter settings of experiments in this section follow implementation details in Sec.~\ref{ssec:implementation}. Table~\ref{tab:stateoftheart} shows the classification accuracy of various methods on these databases, which come from either original or related publications. For some methods, because of the lack of standard deviations in corresponding original or related publications, we list only average classification accuracy. 
In addition, ``$\ast$'' means that we execute the source codes of original papers and obtain corresponding results. Since we encode multi-scale \texttt{BIGD} descriptors using VLAD or IFV in this paper, our proposed methods are represented by ``\texttt{BIGD}+VLAD'' and ``\texttt{BIGD}+IFV'' in Table~\ref{tab:stateoftheart}. 

Table~\ref{tab:stateoftheart_Brodatz} lists the classification accuracy of various methods on the Brodatz database, in which the proposed method (\texttt{BIGD}+IFV) has the best performance. To verify the superiority of the multi-scale sampling strategy, we compare the classification result of the proposed method (\texttt{BIGD}+VLAD) with those in Table~\ref{tab:Brodatz_windowsize}. To fairly compare multi- and single-scale \texttt{BIGD} descriptors, we set their dimensionality to be the same. For example, we select 16 block pairs within a patch in Section~\ref{ssec:implementation}, we select 16 block pairs within a patch; if we have four scales, the number of block pairs at each scale is four; and if we have only one single scale, the number of block pairs at this scale is 16. We notice that \texttt{BIGD}+VLAD with the average classification accuracy of $99.68\%$ outperforms $70\%$ single-scale \texttt{BIGD} descriptors in Table~\ref{tab:Brodatz_windowsize}. In addition, the standard deviation of \texttt{BIGD} + VLAD, $0.14\%$, is smaller than those of all single-scale descriptors in Table~\ref{tab:Brodatz_windowsize}, which supports our claim that compared to the single-scale strategy with the same dimension, the multi-scale strategy has a universal representation and yields robust classification performance. From Table~\ref{tab:KTH-TIPS-2a_windowsize} and Table~\ref{tab:stateoftheart_curet} for KTH-TIPS-2a, we have the same observation about the superiority of the multi-scale strategy. For the CUReT database, the classification accuracy of \texttt{BIGD}+IFV is $0.56\%$ higher than that of  the second best method RP~\cite{liu2012texture} as Table~\ref{tab:stateoftheart_curet} shows.

In Table~\ref{tab:stateoftheart_kthtips}, we compare the classification performance of the proposed method on the KTH-TIPS database with those of other typical and state-of-the-art approaches. The proposed method (\texttt{BIGD}+VLAD) achieves the second best average classification accuracy of $99.02\%$. In contrast, SRP~\cite{liu2015fusing} with the best classification performance on the KTH-TIPS database involves rotation-invariant features. 
Inspired by the randomly sampling strategy in BRIEF~\cite{calonder2010brief}, \texttt{BIGD} achieves comparable performance to SRP because of randomly sampled block pairs, which describe patches at different scales and orientations and enhance the ability of discrimination on rotated textures.

We present the experimental results of various approaches on KTH-TIPS-2a and -2b databases in Tables~\ref{tab:stateoftheart_kthtips2a} and~\ref{tab:stateoftheart_kthtips2b}, respectively. In contrast to the KTH-TIPS database, KTH-TIPS-2a and -2b databases are more challenging because of more image samples and more variations of illuminants. In each class of these two databases, every physical sample corresponds to a set of $108$ images, and testing protocols split each class based on different physical samples rather than the random selection strategy in other databases. Therefore, without any knowledge of testing images, classification tasks on KTH-TIPS-2a and -2b databases become more challenging and suffer the significant decrease of classification accuracy. For the KTH-TIPS-2a database, the proposed method (\texttt{BIGD}+IFV) outperforms DMD~\cite{mehta2016texture} by $0.95\%$. In addition, compared to most single-scale descriptors in Table~\ref{tab:KTH-TIPS-2a_windowsize}, the proposed method (\texttt{BIGD}+VLAD) has the higher classification accuracy and smaller standard deviation, which shows the necessity of the multi-scale sampling strategy. For the KTH-TIPS-2b database, a covariance descriptor, S$_{H}$-SVM~\cite{harandi2014bregman}, achieves the second best accuracy $80.1\%$, which is $2.55\%$ less than the classification accuracy of the proposed method (\texttt{BIGD}+VLAD). From the Table~\ref{tab:stateoftheart}, we observe that the \texttt{BIGD} descriptor has more improvement on more challenging datasets like KTH-TIPS-2a and 2b which supports our claim that it captures the distinctive patterns of patches at different orientations and spatial granularitiy levels.

Though we did an extensive experiments and compared BIGD descriptors with other methods, the DMD descriptor is an important benchmark for comparison for several reasons: (1) DMD is one of the state-of-the-art texture descriptors in texture representation. Unlike most of the earlier work on local texture descriptors, the DMD descriptor does not involve any quantization, thus
retaining the complete information. DMD performs much better than other methods. Also, DMD has dimensionality much lower than Scale Invariant Feature Transform (SIFT) and can be computed using integral image much faster than SIFT shown in~\cite{mehta2016texture}. To show the discriminative ability improvement of our proposed BIGD descriptor through involving the gradient and absolute gradient difference of block pairs in a local patch, we select DMD as an important comparison method but not the only one. (2) DMD has consistent, discriminative texture recognition performance over different datasets and its results are easily reproduced with available source code.

In contrast to the state-of-the-art descriptor DMD, the main contribution of the proposed \texttt{BIGD} descriptor is to involve the gradient and absolute gradient difference of block pairs in a local patch, which improves the distinctiveness of descriptors. To evaluate the benefits of gradient differences on texture classification and guarantee fair comparisons between \texttt{BIGD} and DMD descriptors, the \texttt{BIGD} descriptor in our experiments is designed to have the same dimension as DMD. As shown in Table~\ref{tab:stateoftheart}, if both \texttt{BIGD} and DMD descriptors are encoded by IFV, the former yields $0.12\%\sim5.13\%$ higher classification accuracy than the latter on five databases, where  the highest performance improvement is achieved on the KTH-TIPS-2b database. If we use VLAD as the encoding method, the improvement of classification accuracy on the KTH-TIPS-2b database reaches $6.43\%$. It means that VLAD contributes only $1.30\%$ improvement, which is much less than \texttt{BIGD} descriptors. SIFT~\cite{lowe2004distinctive} as the most typical local descriptor creates a histogram for each key point by partitioning gradient orientations into bins and involving gradient magnitudes as weights. However, quantized orientation features may inevitably result in the loss of information. The proposed \texttt{BIGD} descriptor extracts intensity- and gradient-difference features at multiple orientations without quantization, which retains the discriminative power of features. Last but not the least, the computation efficiency of BIGD is similar to that of DMD, around 100 times faster than SIFT, which has been evaluated in~\cite{cimpoi2014describing} so we do not repeat the details here. Given that $W$ and $H$ denote the width and height of an image and $d$ stands for the feature dimension of a local texture descriptor, $d$ times of local feature difference calculation corresponds to a computational complexity of $O(d)$ for one local descriptor. We repeat this procedure of local feature extraction over an entire image, so the computational complexity of BIGD and DMD are equivalent, i.e. $O(WHd)$.

Deep convolutional network-based features have shown their strong ability as a universal representation in classification or recognition tasks. We list several deep convolutional network-based approaches such as an effective texture descriptor FV-CNN proposed by Cimpoi et al.~\cite{cimpoi2015deep} in Table~\ref{tab:stateoftheart} and borrow their classification results from~\cite{liu2017local}. However, Cimopoi et al's CNN-based methods extracts and train features from color images, while our method only uses gray-scale images but still comparable. And as~\cite{liu2017local} shows, global CNN activations lack geometric invariance resulting in their robustness limitations for recognizing images with high variations. In addition, although deep features are obtained from pre-trained AlexNet, the training process of AlexNet on the ImageNet database requires high computational cost. In contrast, the extraction of \texttt{BIGD} features does not need extra training steps. In addition, our hand-crafted approach is more interpretable. Since gradients are more resilient to photometric changes than intensities, the difference of gradients in \texttt{BIGD} describes the variations of gradients in a local patch and improves distinctiveness. 
Therefore, we still believe our \texttt{BIGD} descriptor is among state-of-the-art, hand-crafted local descriptors.

\section{Discussion on the Rotation and Scale Invariance}
\label{secsec:Analysis for the Rotation and Scale Invariance}
General methods for enabling rotation invariance include three main categories. (1) Dominant orientation estimation: estimating a dominant orientation of each local image patch and then aligning the local patch with respective the orientation or generating a weighted orientation distribution statistics over the orientation; this method suffers from unreliable estimation for cases lacking sharp edges with no dominant orientations or cases including corners with multiple domination orientations; (2) Pattern grouping: rotated versions of the same local structure pattern are grouped into one pattern; one typical example belonging to such category is ``rotation invariant'' LBP (ri-LBP); this method only achieves rotation invariance for short-range local regions while yielding the discrimination ability for long-range structural information; (3) Data augmentation and randomness: adding rotated versions of training samples to the training set during  visual dictionary (i.e. textons) learning or applying random sampling / random projection in the process of local feature extraction. Rotating local texture patches requires more cluster centers and increases the complexity of clustering texture clusters. Our BIGD method belongs to the 3rd category. Therefore, each category has its advantages and disadvantages and should be chosen according to specific applications and requirements such as discrimination ability, and computation efficiency, or robustness. \\

Similarly, we summarize three main approaches for scale invariance. (1) Dominant scale estimation: applying a large number of filter banks to choose dominant scales through maximum filter responses; this method suffers from not accurate estimation especially for the cases without obvious dominant scales in real-world scenarios; (2) Fractal analysis, which provides an alternative with well-founded mathematics; however, fractals lack of discriminative power of different structures or textures; (3) Data augmentation: adding different scales of training samples to the training set during visual words' (i.e. textons) learning or applying a multi-level or multi-resolution strategy in the process of local feature extraction; which is not exactly scale invariant but just brings less sensitiveness to scale variations. Our BIGD method belongs to the 3rd category. In short, the selection of which method for scale invariance still depends on realistic applications and requirements.
\section{Conclusions}
\label{sec:conclusion}

In this paper, we introduced a new local descriptor, block intensity and gradient difference (\texttt{BIGD}), which gains superior distinctiveness compared to state-of-the-art local descriptors while maintaining high computational efficiency. BIGD compared intensity and gradient differences between pairwise blocks within a patch to achieve more robustness for illumination variations and employed a multi-scale random sampling strategy to characterize the structural patterns of the patch at multiple orientations and granularities to achieve more robustness for rotation and scale variations. The extracted BIGD descriptors were then encoded by VLAD or IFV to obtain a discriminative full image representation. The superior performance of our approach was demonstrated by an extensive evaluation on public texture databases. In future work, we will improve the discriminative power of the \texttt{BIGD} descriptor on rotation variations and extend it to other computer vision tasks such as object recognition and tracking tasks. 

\section*{References}

\bibliography{mybibfile}

\end{document}